\documentclass[epj]{svjour}
\RequirePackage{graphicx}

\usepackage[english]{babel}
\usepackage[latin1]{inputenc}
\usepackage{amsmath,amssymb}
\usepackage{bm,graphics,color}

\begin{document}

\title{Traces of Lorentz symmetry breaking in a Hydrogen atom at ground state: }

\author{L.H.C. Borges\inst{1}\thanks{e-mail: luizhenriqueunifei@yahoo.com.br}
        \and
        F.A. Barone\inst{2}\thanks{e-mail: fbarone@unifei.edu.br}
					}

\institute{Universidade Federal do ABC, Centro de Ci\^encias Naturais e Humanas, Rua Santa Ad\'elia, 166, 09210-170, Santo Andr\'e, SP, Brasil.
           \and
           IFQ - Universidade Federal de Itajub\'a, Av. BPS 1303, Pinheirinho, Caixa Postal 50, 37500-903, Itajub\'a, MG, Brazil.
          }

\date{Received: date / Accepted: date}

\abstract{
Some traces of a specific Lorentz symmetry breaking scenario in the ground state of the Hydrogen atom are investigated. It is used standard Rayleigh-Schr\"odinger perturbation theory in order to obtain the corrections to the the ground state energy and wave function. It is shown that an induced four-pole moment arises, due to the Lorentz symmetry breaking. The model considered is the one studied in reference \cite{BBHepjc2014}, where the Lorentz symmetry is broken in the electromagnetic sector.
%
%\PACS{
%      {PACS-key}{discribing text of that key}   \and
%      {PACS-key}{discribing text of that key}
%     } % end of PACS codes
} %end of abstract

\maketitle

%%%%%%%%%
\section{Introduction}
\label{I}
%%%%%%%%%

The Lorentz symmetry breaking is a subject which has been considered in many of contexts nowadays, mainly in what concerns the topics related to the so called Standard Model Extended (SME) \cite{PRD6760,PRD116002}. Among them, it highlights the Lorentz-symmetry breaking signs in atomic and molecular physics. There are two main reasons that motivate this kind of study, namely: the search for physical situations where the Lorentz symmetry breaking could be evinced in our ordinary world and the search for upper bounds imposed on the Lorentz symmetry breaking parameters (a complete list of upper bounds for these parameters can be found in reference \cite{Data}).   

In this context we can stand out, for instance, the CPT and Lorentz symmetry breaking signs on the hydrogen and/or antihydrogen atoms due to modifications in Dirac equation \cite{PRL82,NPB717!86,NPA790!635c,arXiv1506.01706,NIMPRB221!6,arXiv0011272,IJMPA30!6211,arXiv0003223,JMP48!092302,PS72!c38}, the SME effects induced in hydrogen molecules \cite{PRD70!076004}, non minimal coupling effects on hydrogen spectrum \cite{PRD74!065009}, Hydrogen spectrum in models with Lorentz symmetry breaking for dimensional-five models \cite{PRD87!125022}, atomic physics in electromagnetic cavities \cite{PRD67!056006} and maser physics \cite{PRD63!111101R}, the influence of Chern-Simons-type term on the hydrogen atom \cite{arXiv1505.01754} and so on. 

Although we can find some literature considering the hydrogen-like systems in Lorentz-symmetry breaking scenarios \cite{PRD116005}, there are some subjects not yet well explored for such kind of systems. We can cite, for instance, some effects induced on the hydrogen-like atoms by Lorentz symmetry breaking terms in the electromagnetic sector. These effects may occur not only in the spectrum of the system, but we can also obtain interesting results by turning out attention for the probability distribution of the wave functions.  

In this paper we study the effects of a Lorentz symmetry breaking scenario on the ground state of the hydrogen atom. The Lorentz symmetry breaking is taken on the the gauge sector. That is an important subject due two main reasons: the hydrogen atom properties can be measured in the laboratory with high precision and the ground state of the hydrogen atom is spherically symmetric in a Lorentz symmetric model. So any anisotropy found in this state could be an evidence of a possible Lorentz symmetry breaking and it is important to investigate if, nowadays, these effects could be measured by spectroscopic methods. 

In particular, we consider the model studied in reference \cite{BBHepjc2014} where we have a Lorentz symmetry breaking scenario parametrized by just one single background vector and where the Lorentz symmetry is broken only in the electromagnetic sector. We consider the interaction energy between two point-like charges, taking the background vector as a small quantity. In this sense, the interaction between two charges can be written as the Coulomb one added by a small correction. The effects of this correction induces modifications on the ground state of the hydrogen atom, investigated in this paper by using standard Rayleigh-Schr\"odinger perturbation theory. 

Our results suggest that the effects in the ground state of the hydrogen atom are very small and completely out of reach for any spectroscopic experiment nowadays. That is an indication that this kind of study is not a profitable way to search for upper bounds of Lorentz symmetry-breaking parameters for models where this symmetry is broken only in the gauge sector. In spite of this, in the theoretical point of view, it is an interesting subject once it evinces the possible effects on the matter of a Lorentz-symmetry breaking on the gauge sector.

Section (\ref{II}) of this work is devoted to present some general aspects of the model we study. Some care must be taken in redefining the charge of a point particle. In section (\ref{III}) we obtain the corrections to the ground state of the hydrogen atom in lowest order in the background vector. In section (\ref{IV}) we show that there is an induced four-pole electric moment on the ground state of the hydrogen atom. This induced four-pole leads to an interaction between two hydrogen atoms, studied in section (\ref{V}). Section (\ref{concl}) is devoted for our conclusions and final remarks.

%%%%%%%%%
\section{The model}
\label{II}
%%%%%%%%%

In the work of reference \cite{BBHepjc2014} it is considered the model described by the Lagrangian density
%%%%%%%%%
\begin{eqnarray}
\label{lagEm}
{\cal L}=-\frac{1}{4}F_{\mu\nu}F^{\mu\nu}-\frac{1}{2\gamma}\left(\partial_{\mu}A^{\mu}\right)^{2}-\frac{1}{2}v^ {\mu}v_{\nu}F_{\mu\lambda}F^{\nu\lambda}\cr\cr
+J^{\mu}A_{\mu}\ ,
\end{eqnarray}
%%%%%%%%%
where $A^{\mu}$ is the vector field, $F^{\mu\nu}$ its corresponding field strength, $J^{\mu}$ an external source, $\gamma$ is a gauge fixing parameter and $v^{\mu}$ is a background dimensionless vector, taken to be constant and uniform, which accounts for the Lorentz symmetry breaking. The metric is $(+,-,-,-)$ in a $3+1$ spacetime.

The model (\ref{lagEm}) is a special case of the one proposed in the paper \cite{CKPRD116002} where the electromagnetic field couples to a four-rank background tensor $k_{F}$. If this tensor is parameterized by just a single background vector $v^{\mu}$, it is, $(k_{F})_{\alpha\beta\sigma\tau}=1/2(\eta_{\beta\tau}v_{\alpha}v_{\sigma}+\eta_{\alpha\sigma}v_{\beta}v_{\tau}-\eta_{\alpha\tau}v_{\beta}v_{\sigma}-\eta_{\beta\sigma}v_{\alpha}v_{\tau})$, we have the model given in (\ref{lagEm}).

With the results of \cite{BBHepjc2014}, one can show that the interaction energy between two point-like steady opposite charges, $\sigma$ and $-\sigma$, is given by
%%%%%%%%%
\begin{eqnarray}
\label{EnerIEM}
E=-\frac{\sigma^{2}}{4\pi}\frac{\sqrt{1-{\bf v}^{2}}}{1+v^2}\left[{\bf r}^{2}+\frac{({\bf v}\cdot{\bf r})^{2}}{1-{\bf v}^{2}}\right]^{-1/2}\ ,
\end{eqnarray}
%%%%%%%%% 
where ${\bf r}$ is the distance vector from the positive charge to the negative one. It is important to mention that $v^{\mu}$ must be a small quantity.

From now on we shall consider only the quadratic corrections in the background vector $v^{\mu}$ (the lowest order ones).

Defining the electric charge
%%%%%%%%%
\begin{equation}
e=\frac{\sigma}{\sqrt{4\pi}}\ ,
\end{equation}
%%%%%%%%%
expanding the energy (\ref{EnerIEM}) in lowest order in $v^{\mu}$ and taking a coordinates system where ${\bf v}$ lies along the $z$-axis and the positive charge is placed at the origin, we have
%%%%%%%%%
\begin{eqnarray}
\label{EnerIEM2}
E\approx -\frac{e^{2}}{r}+\frac{e^{2}[2(v^{0})^2-{\bf v}^{2}]}{2r}+\frac{e^{2}{\bf{v}}^{2}\cos^{2}\theta}{2r} \ ,
\end{eqnarray}
%%%%%%%%% 
where, $\theta$ is the polar angle in spherical coordinates, which is the angle between the vector $\bf r$ and ${\bf v}$. The first term in (\ref{EnerIEM2}) is the Coulombian interaction and the second term is a correction due the presence of the background vector $v^{\mu}$. In fact, the second term in (\ref{EnerIEM2}) is the lowest correction, in powers of $v^{\mu}$, to the Coulomb interaction.

In a quantum level, the interaction between two opposite charges for the proposed model is governed by the Hamiltonian
%%%%%%%%
\begin{eqnarray}
\label{HIN}
H=H_{0}+\Delta H_{a}+\Delta H_{b} \ ,
\end{eqnarray}
%%%%%%%%%
where
%%%%%%%%
\begin{eqnarray}
\label{HInp}
H_{0}=\frac{P^{2}}{2\mu}-\frac{e^{2}}{r}
\end{eqnarray}
%%%%%%%%% 
is the standard Hamiltonian of a Hydrogen atom (with $P$ standing for the momentum operator and $\mu$, the mass of the particle \cite{Zettili,Cohen,Sakurai}) and 
%%%%%%%%
\begin{eqnarray}
\label{HI}
\Delta H_{a}=\frac{e^{2}{\bf{v}}^{2}\cos^{2}\theta}{2r}\ \ \ \ \ ,\ \ \ \ \ \Delta H_{b}=\frac{e^{2}[2(v^{0})^2-{\bf v}^{2}]}{2r}
\end{eqnarray}
%%%%%%%%% 
are small perturbative stationary corrections, both of the same order (${\bf v}^{2}$).

We are looking for the most relevant contributions introduced by the Lorentz symmetry breaking term of the model (\ref{lagEm}). So we shall use, as the unperturbed problem, the non-relativistic hydrogen atom. It is justified because the relativistic contributions are small in comparison with the non-relativistic ones, and take them into account in this case would lead to Lorentz symmetry breaking corrections to the fine structure, hyperfine structure and to the Lamb-shift.

In spite of this, it is important to highlight that the fine and hyperfine structures and the Lamb-shift are higher contributions in comparison with the ones imposed by the Lorentz symmetry breaking terms.

%%%%%%%%%
\section{Ground state corrections}
\label{III}
%%%%%%%%%

The eigenvalues and the eigenstates of the unperturbed Hamiltonian (\ref{HInp}) are well known from standard treatment of the hydrogen atom \cite{Zettili,Cohen,Sakurai}. Taking the quantum numbers as $n,l,m$ (principal, angular and azimuthal, respectively) and defining the Bohr's radius $a_{0}=\hbar^{2}/(\mu e^{2})$ we have the non-perturbed energy levels
%%%%%%%%%
\begin{equation}
\label{energiasnp}
E^{(0)}_{n}=\frac{E^{(0)}_{1}}{n^{2}}
\end{equation}
%%%%%%%%%
where we defined the ground energy
%%%%%%%%
\begin{eqnarray}
\label{EH}
E^{(0)}_{1}=-\frac{e^{2}}{2a_{0}}\ . 
\end{eqnarray}
%%%%%%%%% 

We shall use standard bra-ket notation, and denote a non-perturbed arbitrary state by $|n,l,m\rangle$ and their corresponding components on the position basis $|{\bf r}\rangle$ by
%%%%%%%%%
\begin{equation}
\label{zxc1}
\langle{\bf r}|n,l,m\rangle=R_{nl}(r)Y_{l}^{m}(\theta,\phi)\ ,
\end{equation}
%%%%%%%%%
with $R_{nl}(r)$ standing for the radial function of the hydrogen atom and $Y_{l}^{m}(\theta,\phi)$, the spherical harmonics \cite{Zettili}.

We shall consider the effects of the correction (\ref{HI}) perturbatively. The ground state is non-degenerated, so its lowest order correction produced by the perturbing hamiltonian is given by
%%%%%%%%
\begin{eqnarray}
\label{CE}
E^{(1)}_{1}=\left\langle 1,0,0\right|(\Delta H_{a}+\Delta H_{b})\left|1,0,0\right\rangle \ .
\end{eqnarray}
%%%%%%%%%

Each contribution in (\ref{CE}) can be calculated by using Eq. (\ref{zxc1}) and the definitions (\ref{HI}),
%%%%%%%%
\begin{eqnarray}
\label{CEI}
\left\langle 1,0,0\right|\Delta H_{a}\left|1,0,0\right\rangle=\frac{e^{2}{\bf v}^{2}}{2}\int^{\infty}_{0}dr \ r \ R_{10}(r)R_{10}(r)\cr\cr
\times\int^{2\pi}_{0}d\phi\int^{\pi}_{0}d\theta\sin\theta\cos^{2}\theta \ [Y^{0}_{0}(\theta,\phi)]^{*}Y^{0}_{0}(\theta,\phi) \ \cr\cr
\left\langle 1,0,0\right|\Delta H_{b}\left|1,0,0\right\rangle=\frac{e^{2}[2(v^{0})^{2}-{\bf v}^{2}]}{2}\times\cr\cr
\times\int^{\infty}_{0}dr \ r \ R_{10}(r)R_{10}(r)\cr\cr
\times\int^{2\pi}_{0}d\phi\int^{\pi}_{0}d\theta\sin\theta \ [Y^{0}_{0}(\theta,\phi)]^{*}Y^{0}_{0}(\theta,\phi)\ .\ \ 
\end{eqnarray}
%%%%%%%%%

The above integrals can be performed by using the fact that
%%%%%%%%
\begin{eqnarray}
\label{EHI}
R_{10}(r)=2a_{0}^{-3/2}\exp(-r/a_{0})\cr\cr
Y^{0}_{0}(\theta,\phi)=\frac{1}{\sqrt{4\pi}}\cr\cr
\int^{\infty}_{0}dr \ r^{n}\exp(-\alpha r)=\frac{n!}{\alpha^{n+1}} \ .
\end{eqnarray}
%%%%%%%%%

Collecting terms, we have the energy for the ground state, $n=1$, of the hydrogen atom in lowest order in the background field
%%%%%%%%
\begin{eqnarray}
\label{EHI2}
E_{1}&=&E^{(0)}_{1}+E^{(1)}_{1}\cr\cr
&=&E^{(0)}_{1}\left(1-\frac{2}{3}[{\bf v}^{2}-3(v^{0})^{2}]\right) \ .
\end{eqnarray}
%%%%%%%%%

It is important to mention that the result (\ref{EHI2}) is in perfect agreement with the corresponding one of reference \cite{PRD116005}. In this reference, the parameter ${\tilde\kappa}_{0}$ (defined just before Eq. (42)), must be calculated with the parameterization that we presented in the paragraph before Eq.(\ref{EnerIEM}). 

For any measurement, the Lorentz symmetry breaking effects must be small and entirely contained in the experimental error nowadays. The experimental value for the ground state energy of the hydrogen atom is $E_{1(e)}^{(0)}=-13.60569253eV$ with an error given by $\delta E_{1(exp)}=3\times10^{-7}eV$ \cite{measure}. Overestimating the energy correction in (\ref{EHI2}), we can say that it is of the same order of $\delta E_{1(exp)}$, so
%%%%%%%%%
\begin{equation} 
\label{estimativa}   
-E^{(0)}_{1}\frac{4}{3}{\bf v}^{2}\cong3\times10^{-7}eV\Rightarrow|{\bf v}|\cong10^{-4}\ .
\end{equation}
%%%%%%%%%

In spite of the value (\ref{estimativa}) to be an overestimation, one can notice that it is a very small quantity. In fact, recent results \cite{Data} stipulate an upper bound smaller than $10^{-19}$ for this kind of Lorentz symmetry breaking parameter, what makes these effects to be out of reach for any spectroscopic measurement nowadays.

The meaning of the result (\ref{EHI2}) is the same one interpreted in the Lorentz invariant case, namely, it is the ionization energy of the hydrogen atom. 
  
The first correction to the lowest energy eigenstate is given by \cite{Sakurai} 
%%%%%%%%
\begin{eqnarray}
\label{WHI}
\left|1,0,0\right\rangle^{1}=\sum_{n\not=1,l,m}\frac{\left\langle n,l,m\right|(\Delta
H_{a}+\Delta
H_{b})\left|1,0,0\right\rangle}{E^{(0)}_{1}-E^{(0)}_{n}}\cr\cr
\times\left|n,l,m\right\rangle .
\end{eqnarray}
%%%%%%%%%

In the coordinates basis, we have
%%%%%%%%
\begin{eqnarray}
\label{WHI2}
\left\langle n,l,m\right|\Delta
H_{a}\left|1,0,0\right\rangle=\cr\cr
=\frac{e^{2}{\bf v}^{2}}{2}\int^{\infty}_{0}\!\!dr\ rR_{nl}(r)R_{10}(r)\cr\cr
\times\int^{2\pi}_{0}\!\!d\phi\int^{\pi}_{0}d\theta\sin\theta\cos^{2}\theta[Y^{m}_{l}(\theta,\phi)]^{*}Y^{0}_{0}(\theta,\phi)\cr\cr
\left\langle n,l,m\right|\Delta
H_{b}\left|1,0,0\right\rangle=\cr\cr
=\frac{e^{2}[2(v^{0})^{2}-{\bf v}^{2}]}{2}\int^{\infty}_{0}\!\!dr\ rR_{nl}(r)R_{10}(r)\cr\cr
\times\int^{2\pi}_{0}\!\!d\phi\int^{\pi}_{0}d\theta\sin\theta[Y^{m}_{l}(\theta,\phi)]^{*}Y^{0}_{0}(\theta,\phi).\ 
\end{eqnarray}
%%%%%%%%%

The normalized spherical harmonics can be written as \cite{Arfken}
%%%%%%%
\begin{eqnarray}
\label{SH}
Y^{m}_{l}(\theta,\phi)=(-1)^{m}\sqrt{\left(\frac{2l+1}{4\pi}\right)\frac{(l-m)!}{(l+m)!}}\cr\cr
\times P^{m}_{l}(\cos\theta)\exp(im\phi) \ ,
\end{eqnarray}
%%%%%%%%%
where $P^{m}_{l}(\cos\theta)$ are the associated Legendre functions \cite{Arfken}. Substituting (\ref{SH}) in (\ref{WHI2}) and using the fact that
%%%%%%%%%
\begin{equation}
\int^{2\pi}_{0}d\phi\exp(-im\phi)=2\pi\delta_{m0}
\end{equation}
%%%%%%%%%
(where $\delta_{m0}$ is the Kronecker delta), it is possible rewrite (\ref{WHI2}) in the following form
%%%%%%%%%
\begin{eqnarray}
\label{WHI3}
\left\langle n,l,m\right|(\Delta
H_{a}+\Delta
H_{b})\left|1,0,0\right\rangle=\cr\cr
=\frac{e^{2}}{4}\delta_{m0}\sqrt{2l+1}\int^{\infty}_{0}dr \ rR_{nl}(r)R_{10}(r)\cr\cr
\times\int^{\pi}_{0}d\theta \ P^{0}_{l}(\cos\theta)\sin\theta\Bigl({\bf v}^{2}\cos^{2}\theta+[2(v^{0})^2-{\bf v}^{2}]\Bigr) .\ 
\end{eqnarray}
%%%%%%%%%

Using the results \cite{Arfken}
%%%%%%%
\begin{eqnarray}
\label{RL}
\int^{\pi}_{0}d\theta P^{m}_{p}(\cos\theta)P^{m}_{q}(\cos\theta)\sin\theta=\frac{2\delta_{pq}}{2q+1}\frac{(q+m)!}{(q-m)!}\cr\cr
\cos^{2}\theta=\frac{2}{3}P^{0}_{2}(\cos\theta)+\frac{1}{3}P^{0}_{0}(\cos\theta)\ \ , \ \ 1=P^{0}_{0}(\cos\theta) ,\ \ \ 
\end{eqnarray}
%%%%%%%%%
we can show that
%%%%%%%
\begin{eqnarray}
\label{LI}
\int^{\pi}_{0}d\theta \ P^{0}_{l}(\cos\theta)\sin\theta\Bigl({\bf v}^{2}\cos^{2}\theta+[2(v^{0})^2-{\bf v}^{2}]\Bigr)=\cr\cr
={\bf v}^{2}\Biggl(\frac{4}{15}\delta_{l2}+\frac{2}{3}\delta_{l0}\Biggr)+[2(v^{0})^2-{\bf v}^{2}]2\delta_{l0} .\cr
\ 
\end{eqnarray}
%%%%%%%%%

With the aid of (\ref{LI}) and (\ref{WHI3}), we can rewrite Eq. (\ref{WHI}) in the form
%%%%%%%
\begin{eqnarray}
\label{WHI5}
\left|1,0,0\right\rangle^{1}=\frac{e^{2}}{3}\sum_{n\not=1}\frac{1}{E^{(0)}_{1}-E^{(0)}_{n}}\cr\cr
\times\Biggl[\frac{{\bf v}^{2}}{\sqrt{5}}\left(\int^{\infty}_{0}dr \ rR_{n2}(r)R_{10}(r)\right)\left|n,2,0\right\rangle\cr\cr
+[3(v^{0})^{2}-{\bf v}^{2}]\left(\int^{\infty}_{0}dr \ rR_{n0}(r)R_{10}(r)\right)\left|n,0,0\right\rangle\Biggr]\ .
\end{eqnarray}
%%%%%%%%%

In order to calculate the radial integrals in (\ref{WHI5}), we must use the expansions \cite{Arfken}
%%%%%%%
\begin{eqnarray}
\label{RFH}
R_{nl}(r)=\left[\frac{\alpha^{3}(n-l-1)!}{2n(n+l)!}\right]^{1/2}\cr\cr
\times\exp\left(-\frac{\alpha r}{2}\right)\left(\alpha r\right)^{l}L^{2l+1}_{n-l-1}\left(\alpha r\right) \ ,
\end{eqnarray}
%%%%%%%%% 
where $\alpha=2/na_{0}$ and $L^{2l+1}_{n-l-1}\left(\alpha r\right)$ are the associated Laguerre functions \cite{Arfken}, which can be written as a series \cite{Arfken},
%%%%%%%
\begin{eqnarray}
\label{LGF}
L^{k}_{q}\left(x\right)=\sum^{q}_{s=0}(-1)^{s}\frac{(q+k)!}{(q-s)!(k+s)!s!}x^{s} \ ; \ k>-1\ .
\end{eqnarray}
%%%%%%%%% 

Substituting Eq's (\ref{RFH}) and (\ref{LGF}) in the first radial integral inside brackets in Eq. (\ref{WHI5}) and using the explicit form of $R_{10}(r)$, shown in the first Eq. (\ref{EHI}), we obtain
%%%%%%%
\begin{eqnarray}
\label{RIEI}
\int^{\infty}_{0}dr\ rR_{n2}(r)R_{10}(r)=\frac{16}{a_{0}}\frac{[(n-3)!(n+2)!]^{1/2}}{(n+1)^{4}}\cr\cr
\times\sum^{n-3}_{t=0}\frac{(-1)^{t}2^{t}}{t!(n-3-t)!(t+5)(t+4)(n+1)^{t}}\ ; \ n\geq3 ,
\end{eqnarray}
%%%%%%%%% 
where we have used the integral in Eq. (\ref{EHI}). Proceeding in a similar way, we have
%%%%%%%%%
\begin{eqnarray}
\label{RIEI2}
\int^{\infty}_{0}dr \ r \ R_{n0}(r)R_{10}(r)=\frac{4}{a_{0}}\left[\frac{n^{-1/2}n!}{(n+1)^{2}}\right]\cr\cr
\times\sum^{n-1}_{s=0}\frac{(-1)^{s}2^{s}}{s!(n-1-s)!(n+1)^{s}} \ ; \ n\geq2 \ .
\end{eqnarray}
%%%%%%%%%

Substituting Eq's (\ref{RIEI}) and (\ref{RIEI2}) in (\ref{WHI5}) and using the fact that $1/E^{(0)}_{1}-E^{(0)}_{n\not=1}=2a_{0}n^{2}/e^{2}(1-n^{2})$ we obtain the lowest order correction to the ground state energy,
%%%%%%%
\begin{eqnarray}
\label{WHI6}
\left|1,0,0\right\rangle^{1}=\frac{8}{3}\Biggl[\frac{4{\bf v}^{2}}{\sqrt{5}}\sum^{\infty}_{n=3}\sum^{n-3}_{t=0}C_{1}\left(n,t\right)\left|n,2,0\right\rangle\cr\cr
+[3(v^{0})^{2}-{\bf v}^{2}]\sum^{\infty}_{n=2}\sum^{n-1}_{s=0}C_{2}\left(n,s\right)\left|n,0,0\right\rangle\Biggr] \ ,
\end{eqnarray}
%%%%%%%%%
where
%%%%%%%
\begin{eqnarray}
\label{C1C2}
C_{1}\left(n,t\right)&=&\frac{(-1)^{t}2^{t}n^{2}[(n-3)!(n+2)!]^{1/2}}{(n+1)^{4}(1-n^{2})(n+1)^{t}}\cr\cr
&\ &\times\frac{1}{(t+5)(t+4)(n-3-t)!t!}\ ;\cr\cr
C_{2}\left(n,s\right)&=&\frac{(-1)^{s}2^{s}n^{3/2}n!}{(n+1)^{2}(1-n^{2})(n+1)^{s}}\cr\cr
&\ &\times\frac{1}{(n-1-s)!s!} \ .
\end{eqnarray}
%%%%%%%%%

At this time, some points are in order. The series which involve $C_{1}\left(n,t\right)$ and $C_{2}\left(n,s\right)$ in (\ref{WHI6}) are convergent. The non-perturbed ground state is spherically symmetric and involves only the quantum numbers $n=1,\ l=0,\ m=0$. The perturbation in the ground state involves the values $n=2,3,4,5,...$ for the principal quantum number, $l=0,2$ for the angular-momentum quantum number and only $m=0$ for the azimuthal quantum number.

In summary, in first order, the ground state is given by
%%%%%%%%%
\begin{eqnarray}
\label{EFUN}
\left|{\bf v}\right\rangle=\left|1,0,0\right\rangle+\left|1,0,0\right\rangle^{1} \ ,
\end{eqnarray}
%%%%%%%%% 
with $\left|1,0,0\right\rangle^{1}$ given by (\ref{WHI6}).

%%%%%%%%%
\section{Induced electric dipole and four-pole and magnetic dipole}
\label{IV}
%%%%%%%%%

In order to investigate the anisotropies, imposed by the Lorentz symmetry breaking, on the ground state of the hydrogen atom, let us check if there is an induced dipole or four-pole moments in this state. 

The dipole moment operator is defined by \cite{Cohen} ${\bf{D}}=e{\bf{R}}$, where we have the position operator, ${\bf R}=R_{1}{\hat x}+R_{2}{\hat y}+R_{3}{\hat z}$ with $R_{1}$, $R_{2}$ and $R_{3}$ standing for the Cartesian coordinates operator for $x$, $y$ and $z$, respectively. Using Eq's (\ref{WHI6}) and (\ref{EFUN}) we can write
%%%%%%%%%
\begin{eqnarray}
\label{zxc2}
\langle{\bf v}|{\bf D}|{\bf v}\rangle=\langle1,0,0|e{\bf R}|1,0,0\rangle\cr\cr
+\ ^{1}\!\langle1,0,0|e{\bf R}|1,0,0\rangle+\langle1,0,0|e{\bf R}|1,0,0\rangle^{1}+{\cal O}({\bf v}^{4}) .
\end{eqnarray}
%%%%%%%%%

Using the fact that
%%%%%%%%%
\begin{eqnarray}
\label{zxc4}
Y_{0}^{0}=\frac{1}{\sqrt{4\pi}}\ \ ,\ \ Y_{2}^{0}(\theta,\phi)=\sqrt{\frac{5}{16\pi}}(3\cos^{2}\theta-1)\cr\cr
x=r\sin\theta\cos\phi\ \ ,\ \ y=r\sin\theta\sin\phi\ \ ,\ \ z=r\cos\theta
\end{eqnarray}
%%%%%%%%%
we can show that, up to order ${\bf v}^{2}$, the right hand side of Eq. (\ref{zxc2}) vanishes. So $\langle1,0,0|{\bf D}|1,0,0\rangle=0$ and there is no induced dipole moment on the ground state of the hydrogen atom in order ${\bf v}^{2}$.

The four-pole operator is given by 
%%%%%%%
\begin{eqnarray}
\label{QMO}
Q_{ij}=e\left(3R_{i}R_{j}-{{\bf R}^{2}}\delta_{ij}\right) \ ,
\end{eqnarray}
%%%%%%%%%

For convenience, we define
%%%%%%%
\begin{eqnarray}
\label{DMO3}
\left|\Delta_{2,0}\right\rangle=\frac{32}{3\sqrt{5}}{\bf{v}}^{2}\sum^{\infty}_{n=3}\sum^{n-3}_{t=0}C_{1}\left(n,t\right)\left|n,2,0\right\rangle\cr\cr \left|\Delta_{0,0}\right\rangle=\frac{8}{3}[3(v^{0})^{2}-{\bf v}^{2}]\sum^{\infty}_{n=2}\sum^{n-1}_{s=0}C_{2}\left(n,s\right)\left|n,0,0\right\rangle \ ,
\end{eqnarray}
%%%%%%%%%
in such a way to write Eq. (\ref{WHI6}) in the form
%%%%%%%%%
\begin{equation}
\label{zxc3}
\left|1,0,0\right\rangle^{1}=\left|\Delta_{2,0}\right\rangle+\left|\Delta_{0,0}\right\rangle\ .
\end{equation}
%%%%%%%%%

The expected value of the four-pole components on the corrected ground state (\ref{EFUN}) is given by
%%%%%%%
\begin{eqnarray}
\label{QMO2}
\left\langle {\bf v}\right|Q_{ij}\left|{\bf v}\right\rangle=\left\langle 1,0,0\right|Q_{ij}\left|1,0,0\right\rangle\cr\cr
+\left\langle 1,0,0\right|Q_{ij}\left|\Delta_{2,0}\right\rangle+\left\langle 1,0,0\right|Q_{ij}\left|\Delta_{0,0}\right\rangle\cr\cr
+\left\langle \Delta_{2,0}\right|Q_{ij}\left|1,0,0\right\rangle+\left\langle \Delta_{0,0}\right|Q_{ij}\left|1,0,0\right\rangle \ ,
\end{eqnarray}
%%%%%%%%%
where we neglected terms in order ${\bf v}^{4}$.

Using Eq's (\ref{zxc4}), (\ref{QMO}), (\ref{DMO3}) and (\ref{zxc3}) we can show that
%%%%%%%
\begin{eqnarray}
\label{QMOF}
\left\langle 1,0,0\right|Q_{ij}\left|1,0,0\right\rangle=\left\langle 1,0,0\right|Q_{ij}\left|\Delta_{0,0}\right\rangle\cr\cr
=\left\langle \Delta_{0,0}\right|Q_{ij}\left|1,0,0\right\rangle=0 \ ,
\end{eqnarray}
%%%%%%%%%
%%%%%%%%%
\begin{eqnarray}
\label{QMOF2}
\left\langle 1,0,0\right|Q_{ij}\left|\Delta_{2,0}\right\rangle=\left\langle \Delta_{2,0}\right|Q_{ij}\left|1,0,0\right\rangle=0\  ;\ i\not=j\cr
\   
\end{eqnarray}
%%%%%%%%%
and
%%%%%%%
\begin{eqnarray}
\label{QMOF3}
\left\langle 1,0,0\right|Q_{33}\left|\Delta_{2,0}\right\rangle=\frac{64}{15}q{{\bf v}^{2}}\cr\cr
\times\sum^{\infty}_{n=3}\sum^{n-3}_{t=0}C_{1}\left(n,t\right)\left(\int^{\infty}_{0}dr\ r^{4}\ R_{10}\left(r\right)R_{n2}\left(r\right)\right) \ .
\end{eqnarray}
%%%%%%%%%

The above integral can be computed following similar steps we have done to obtain (\ref{RIEI}), 
%%%%%%%
\begin{eqnarray}
\label{QMOF5}
\int^{\infty}_{0}dr\ r^{4}\ R_{10}\left(r\right)R_{n2}\left(r\right)=16{a_{0}}^{2}[(n-3)!(n+2)!]^{1/2}\cr\cr
\frac{n^{3}}{(n+1)^{7}}\sum^{n-3}_{w=0}\frac{(-1)^{w}2^{w}(w+6)}{w!(n-3-w)!(n+1)^{w}} \ ; \ w\geq3 \ .
\end{eqnarray}
%%%%%%%%%

For convenience, we define
%%%%%%%%%
\begin{eqnarray}
\label{QMOF7}
C_{3}\left(n,w\right)=\frac{(-1)^{w}2^{w}(w+6)n^{3}[(n-3)!(n+2)!]^{1/2}}{(n+1)^{w+7}(n-3-w)!w!} .
\end{eqnarray}
%%%%%%%%%

Substituting (\ref{QMOF5}) in (\ref{QMOF3}) and using (\ref{QMOF7}), we obtain
%%%%%%%%%
\begin{eqnarray}
\label{QMOF6}
\left\langle 1,0,0\right|Q_{33}\left|\Delta_{2,0}\right\rangle=\frac{1024{a_{0}}^{2}}{15}e{{\bf v}^{2}}\cr\cr
\times\sum^{\infty}_{n=3}\left(\sum^{n-3}_{t=0}C_{1}\left(n,t\right)\right)\left(\sum^{n-3}_{w=0}C_{3}\left(n,w\right)\right) \ .
\end{eqnarray}
%%%%%%%%% 

It is worth mentioning that the sum involving $C_{3}\left(n,w\right)$ in (\ref{QMOF6}) is convergent, and expression (\ref{QMOF6}) can be calculated numerically, the result is
%%%%%%%%%
\begin{equation}
\label{numD}
\left\langle 1,0,0\right|Q_{33}\left|\Delta_{2,0}\right\rangle\cong2.83\times10^{-2}e{\bf v}^{2}a_{0}^{2}
\end{equation}
%%%%%%%%%

In the same manner, we can show that
%%%%%%%%%
\begin{eqnarray}
\label{QMOF8}
\left\langle 1,0,0\right|Q_{11}\left|\Delta_{2,0}\right\rangle=\left\langle 1,0,0\right|Q_{22}\left|\Delta_{2,0}\right\rangle\cr\cr
=-{\frac{1}{2}}\left\langle 1,0,0\right|Q_{33}\left|\Delta_{2,0}\right\rangle\ .
\end{eqnarray}
%%%%%%%%%
and
%%%%%%%%%
\begin{eqnarray}
\label{QMOF9}
\left\langle \Delta_{2,0}\right|Q_{33}\left|1,0,0\right\rangle&=&\left\langle 1,0,0\right|Q_{33}\left|\Delta_{2,0}\right\rangle\nonumber \ ; \\
\left\langle \Delta_{2,0}\right|Q_{11}\left|1,0,0\right\rangle&=&\left\langle 1,0,0\right|Q_{11}\left|\Delta_{2,0}\right\rangle\nonumber \ ; \\ 
\left\langle \Delta_{2,0}\right|Q_{22}\left|1,0,0\right\rangle&=&\left\langle 1,0,0\right|Q_{22}\left|\Delta_{2,0}\right\rangle \ . 
\end{eqnarray}
%%%%%%%%%

Substituting (\ref{QMOF}), (\ref{QMOF2}), (\ref{QMOF8}), (\ref{QMOF9}) in (\ref{QMO2}) we find that the matrix which gives the components of expected value of quadrupole moment operator is given by
%%%%%%%%%%%%%
\begin{eqnarray}
\label{matri1}
\left\langle{\bf v}\right|{\bf{Q}}\left|{\bf v}\right\rangle=
-\left\langle 1,0,0\right|Q_{33}\left|\Delta_{2,0}\right\rangle\bordermatrix{&\cr
							& 1 \ \ &0 \ \ &0 \   \cr
              & 0 \ \ &1 \ \ &0 \       \cr
              & 0 \ \ &0 \ \ &-2 \   \cr} \cr
=-2,83\times10^{-2}e{\bf v}^{2}a_{0}^{2}\bordermatrix{&\cr
							& 1 \ \ &0 \ \ &0 \   \cr
              & 0 \ \ &1 \ \ &0 \       \cr
              & 0 \ \ &0 \ \ &-2 \   \cr} ,
\end{eqnarray}
%%%%%%%%%%%%%
where $\left\langle 1,0,0\right|Q_{33}\left|\Delta_{2,0}\right\rangle$ is given by (\ref{QMOF6}) or, numerically, by (\ref{numD}).

So, due to the Lorentz symmetry breaking, the ground state of the hydrogen atom has it spherical symmetry broken and exhibits an electric four-pole moment. 

The four-pole tensor (\ref{matri1}) was calculated in a reference frame where the background vector is constant and uniform. This statement is valid only for a given inertial frame, what is not the case of a laboratory on Earth. To put these results in an experimental context, one would calculate this four-pole tensor in the laboratory frame, taking into account the fact that the background vector rotates with respect to the laboratory frame as a function of time \cite{arXiv1506.01706}. It would produce a rotation of the induced four-pole calculated above. This rotation does not modify the magnitude of the four-pole tensor whose effects are very small.

The magnetic dipole operator is given
%%%%%%%%%
\begin{equation}
{\bf M}=\frac{1}{2}\frac{e}{m_{e}}{\bf L}
\end{equation}
%%%%%%%%%
where ${\bf L}$ is the orbital angular momentum and $m_{e}$ is the mass of the electron.

Following the same arguments, we can show that the induced magnetic dipole on the atom is given by
%%%%%%%%%
\begin{eqnarray}
\label{bnm1}
\langle{\bf v}|{\bf M}|{\bf v}\rangle=\frac{1}{2}\frac{e}{m_{e}}\Big(\langle1,0,0|{\bf L}|1,0,0\rangle+\cr\cr
+\ ^{1}\!\langle1,0,0|{\bf L}|1,0,0\rangle+\langle1,0,0|{\bf L}|1,0,0\rangle^{1}\Big)+{\cal O}({\bf v}^{4}) .
\end{eqnarray}
%%%%%%%%%

Using Eq. (\ref{WHI6}) and the fact that 
%%%%%%%%%
\begin{eqnarray}
\langle1,0,0|{\bf L}|n,0,0\rangle=\langle1,0,0|{\bf L}|n,2,0\rangle=0\ ,
\end{eqnarray}
%%%%%%%%%
we can show that the induced magnetic dipole (\ref{bnm1}) vanishes up to order ${\bf v}^{4}$, that is
%%%%%%%%%
\begin{equation}
\langle{\bf v}|{\bf M}|{\bf v}\rangle\cong0\ .
\end{equation}
%%%%%%%%%

%%%%%%%%%
\section{Atomic interaction}
\label{V}
%%%%%%%%%

We have seen that the background vector induces an electric four-pole moment on the hydrogen atom in the ground state. As an immediate consequence, we have an interaction between two hydrogen atoms when they are in the ground state.

The interaction between two hydrogen atoms in the ground state is not a novel phenomenon. It is well known that they can interact via Van der Waals forces, but in our case, this interaction is different by two main reasons: it is a four-pole interaction and is induced by the background vector, an external agent. The Van der Waals interaction is a dipole-type interaction induced by the atoms one another.

In this section we study the interaction between two hydrogen atoms, when they are in the ground state, induced by the Lorentz-symmetry breaking in lowest order in the background field.

From the expression for the interaction energy between two electric four-poles, we would obtain an energy in order ${\bf v}^{4}$, that is of superior order in the background vector. So we must investigate how the energy of each atom is modified by the electric field produced by the other atom.

Let us consider two atomic nuclei placed a distance ${\bf r}$ apart, in a coordinate system with the first nucleus placed at the origin and the second one placed at position ${\bf r}$. For the second atom, the position of the electron is taken as ${\bf r}+{\bf R}$, so it is placed a distance ${\bf R}$ apart from its nucleus. We shall restrict ourselves to the case where the distance between the atoms is much higher than the atomic nuclei. This condition is attained by the restriction ${\bf R}^{2}<<{\bf r}^{2}$.

We shall calculate the energy shift of the second atom due to the electric four-pole induced in the first atom $\Delta E_{(2)}^{Q}$. By symmetry, it is equal to the energy shift induced in the first atom by the electric four-pole induced in the second atom. So, the total shift in the energy of the whole system is $\Delta E^{Q}=2\Delta E_{(2)}^{Q}$.

The Hamiltonian of the electron in the second atom is composed by the non-perturbed hydrogen one (\ref{HInp}), a correction like the one in (\ref{HI}) and a four-pole term produced by the first atom, all of them written as functions of ${\bf R}$. We have also the terms which accounts for the interaction between the electron and the nucleus of the first atom with the electron and the nucleus of the second atom. These last terms lead to the Van der Waals interactions, which are well known in the literature. In this paper we shall focus only in the corrections due to the Lorentz symmetry breaking. The contributions which come from (\ref{HInp}) and the correction (\ref{HI}) where taken into account in the previous sections.

So, let us consider the four-pole contribution \cite{Jackson}
%%%%%%%%%%%%%
\begin{eqnarray}
\label{2HYA4} 
\Delta H^{Q}_{(2)}&=&\frac{e}{2|{{\bf R}+{\bf r}}|^{5}}\sum^{3}_{i,j=1}\left({\bf R}+{\bf r}\right)_{i}\left({\bf R}+{\bf r}\right)_{j}Q_{ij}\cr\cr
&
&-\frac{e}{2|{{\bf r}}|^{5}}\sum^{3}_{i,j=1}{\bf r}_{i}{\bf r}_{j}Q_{ij}\cr\cr
&=&\frac{e}{2|{{\bf R}+{\bf r}}|^{5}}\sum^{3}_{i,j=1}\left(R_{i}+r_{i}\right)\left(R_{j}+r_{j}\right)Q_{ij}\cr\cr
&
&-\frac{e}{2|{{\bf r}}|^{5}}\sum^{3}_{i,j=1} r_{i}r_{j}Q_{ij}\ .
\end{eqnarray}
%%%%%%%%%%%%%

In the last line of Eq. (\ref{2HYA4}) we have the interaction between the four-pole induced in the first atom and the nucleus of the second atom. In the third line, we have the interaction between the four-pole of the first atom and the electron of the second atom.

Using Eq. (\ref{matri1}), defining
 %%%%%%%%%%%%%
\begin{equation}
\label{2HYA7} 
Q=\left\langle 1,0,0\right|Q_{33}\left|\Delta_{2,0}\right\rangle\ ,
\end{equation}
%%%%%%%%%%%%%
and keeping only the terms up to second order in $\frac{|{\bf R}|}{|{\bf r}|}$ (because $\frac{|{\bf R}|}{|{\bf r}|}<<1$), we can write Eq. (\ref{2HYA4}) in the form
%%%%%%%%%%%%%
\begin{eqnarray}
\label{2HYA9}
\Delta H^{Q}_{(2)}&\cong &-\frac{15Qe}{4 |{\bf r}|^{7}}\Biggl\{-\frac{1}{5}\left[|{\bf r}|^{2}-5\left({{\bf r}\cdot{\hat{z}}}\right)^{2}\right]\Bigl[|{\bf R}|^{2}\cr\cr
&
&+2\left({{\bf r}\cdot{\bf R}}\right)\Bigr]+\left[1-\frac{7\left({{\bf r}\cdot{\hat{z}}}\right)^{2}}{|{\bf r}|^{2}}\right]\left({{\bf r}\cdot{\bf R}}\right)^{2}\cr\cr
&
&+\left[-\frac{2}{5}|{\bf r}|^{2}+|{\bf R}|^{2}+2\left({{\bf r}\cdot{\bf R}}\right)-\frac{7\left({{\bf r}\cdot{\bf R}}\right)^{2}}{|{\bf r}|^{2}}\right]\cr\cr
&
&\times\left[2\left({{\bf r}\cdot{\hat{z}}}\right)\left({{\bf R}\cdot{\hat{z}}}\right)+\left({{\bf R}\cdot{\hat{z}}}\right)^{2}\right]\Biggr\} \ .
\end{eqnarray}
%%%%%%%%%%%%%

Now we use Rayleigh-Schrödinger perturbation theory in first order to write the correction to the energy of the second atom due to the four-pole Hamiltonian (\ref{2HYA4})  
%%%%%%%%
\begin{eqnarray}
\label{2HYA11}
\Delta E^{Q}_{(2)}=\left\langle{\bf v}\right|\Delta H^{Q}_{(2)}\left|{\bf v}\right\rangle\cong\left\langle 1,0,0\right|\Delta H^{Q}_{(2)}\left|1,0,0\right\rangle\ ,
\end{eqnarray}
%%%%%%%%%
where, in the second line, we used Eq. (\ref{EFUN}) and discarded the contributions which come from $\left|1,0,0\right\rangle^{1}$, once they are of superior order in the background vector. 

Substituting Eq. (\ref{2HYA9}) in (\ref{2HYA11}), using the fact that
%%%%%%%%
\begin{eqnarray}
\label{2HYA15} 
\left\langle 1,0,0\right|{{\bf r}\cdot{\bf R}}\left|1,0,0\right\rangle=\left\langle 1,0,0\right|\left({{\bf R}\cdot{\hat{z}}}\right)\left|1,0,0\right\rangle=&0&\cr\cr
\left\langle 1,0,0\right|\left({{\bf r}\cdot{\bf R}}\right)^{2}\left|1,0,0\right\rangle=&|{\bf r}|^{2}a_{0}^{2}&\cr\cr
\left\langle 1,0,0\right||{\bf R}|^{2}\left|1,0,0\right\rangle=3\left\langle 1,0,0\right|\left({{\bf R}\cdot{\hat{z}}}\right)^{2}\left|1,0,0\right\rangle=&3a_{0}^{2}&\cr\cr
\left\langle 1,0,0\right|\left({{\bf R}\cdot{\hat{z}}}\right)\left({{\bf r}\cdot{\bf R}}\right)\left|1,0,0\right\rangle=\left({{\bf r}\cdot{\hat{z}}}\right)\!\!\!\!\!&a_{0}^{2}&\cr\cr
\left\langle 1,0,0\right||{\bf R}|^{2}\left({{\bf R}\cdot{\hat{z}}}\right)^{2}\left|1,0,0\right\rangle=&\frac{15}{2}a_{0}^{4}&\cr\cr
\left\langle 1,0,0\right||{\bf R}|^{2}\left({{\bf R}\cdot{\hat{z}}}\right)\left|1,0,0\right\rangle=&0&\cr\cr
\left\langle 1,0,0\right|\left({{\bf R}\cdot{\hat{z}}}\right)\left({{\bf r}\cdot{\bf R}}\right)^{2}\left|1,0,0\right\rangle=&0&\cr\cr
\left\langle 1,0,0\right|\left({{\bf R}\cdot{\hat{z}}}\right)^{2}\left({{\bf r}\cdot{\bf R}}\right)\left|1,0,0\right\rangle=&0&\cr\cr
\left\langle 1,0,0\right|\left({{\bf R}\cdot{\hat{z}}}\right)^{2}\left({{\bf r}\cdot{\bf R}}\right)^{2}\left|1,0,0\right\rangle=&\ &\cr
=\frac{3a_{0}^{4}}{2}\Bigl[|{\bf r}|^{2}+2\left({{\bf r}\cdot{\hat{z}}}\right)^{2}\Bigr],
\end{eqnarray}
%%%%%%%%%
$\Delta E^{Q}=2\Delta E_{(2)}^{Q}$ and ${\hat v}={\hat z}$ in the coordinates system we have chosen, we can write the energy shift of the two atoms system due to the electric four-pole interaction
%%%%%%%%
\begin{eqnarray}
\label{Eint2H}
\Delta E^{Q}=\frac{45Qea_{0}^{4}}{2|{\bf r}|^{9}}\left[|{\bf r}|^{2}+7\left({{\bf r}\cdot{\hat v}}\right)^{2}\right]\cr\cr
=\frac{45Qea_{0}^{4}}{2|{\bf r}|^{7}}\left[1+7\cos^{2}(\theta)\right] \ .
\end{eqnarray}
%%%%%%%%%

From the energy (\ref{Eint2H}) we obtain an interaction force between the atoms 
%%%%%%%%
\begin{eqnarray}
\label{2HYA18}
{\bf F}^{Q}&=&-{\bf \nabla}\left(\Delta E^{Q}\right)\cr\cr
&=&\frac{315Qea_{0}^{4}}{2|{\bf r}|^{9}}\left\{\left[1-\frac{9\left({{\bf r}\cdot{\hat v}}\right)^{2}}{|{\bf r}|^{2}}\right]{\bf r}-2\left({{\bf r}\cdot{\hat v}}\right)\hat{v}\right\}\cr\cr
&=&\frac{315Qea_{0}^{4}}{2|{\bf r}|^{8}}\left\{\left[1-9\cos^{2}(\theta)\right]{\hat r}-2\cos(\theta)\hat{v}\right\}
\ ,
\end{eqnarray}
%%%%%%%%%
and a torque on the two atoms system
%%%%%%%%%
\begin{eqnarray}
\label{torque}
\tau^{Q}=-\frac{\partial\Delta E^{Q}}{\partial \theta}=\frac{315Qea_{0}^{4}}{2|{\bf r}|^{7}}\sin(2\theta) \ .
\end{eqnarray}
%%%%%%%%%

The force (\ref{2HYA18}) falls with $r^{8}$. It exhibits a radial component as well as a component along the background vector. It is interesting to notice that, depending on the angle $\theta$, the force (\ref{2HYA18}) can be repulsive or attractive. This fact can be seen if on considers the components of (\ref{2HYA18}) in cylindrical coordinates ($\rho=|{\bf r}|\sin\theta$),
%%%%%%%%%
\begin{eqnarray}
F_{\rho}=\frac{315Qea_{0}^{4}}{2|{\bf r}|^{8}}f_{\rho}(\theta)\ \ \ ,\ \ \ 
F_{z}=\frac{315Qea_{0}^{4}}{2|{\bf r}|^{8}}f_{z}(\theta)
\end{eqnarray}
%%%%%%%%%
where we defined the functions
%%%%%%%%%
\begin{eqnarray}
f_{\rho}(\theta)=\left[1-9\cos^{2}(\theta)\right]\sin(\theta)\cr\cr
f_{z}(\theta)=-\left[1+9\cos^{2}(\theta)\right]\cos(\theta)\ ,
\end{eqnarray}
%%%%%%%%%
which control the dependence of the signals of $F_{\rho}$ and $F_{z}$ with $\theta$.

\begin{figure}[!ht]
 \centering
   \includegraphics[scale=0.20]{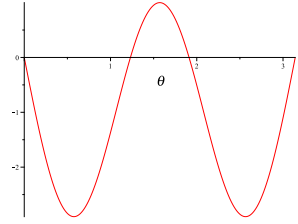}
   \caption{Plot for $f_{\rho}(\theta)$.}
  \label{g1}
\end{figure}

\begin{figure}[!ht]
 \centering
   \includegraphics[scale=0.20]{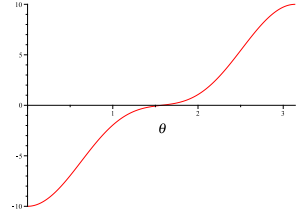}
   \caption{Plot for $f_{z}(\theta)$.}
  \label{g2}
\end{figure}

In figures (\ref{g1}) and (\ref{g2}) we have, respectively, plots for $f_{\rho}$ and $f_{z}$. In the intervals $\theta=[0,\arccos(1/3)]$ and $\theta=[\pi-\arccos(1/3),\pi]$ the function $f_{\rho}$ is negative. For $\theta=[\arccos(1/3),\pi-\arccos(1/3)]$, $f_{\rho}$ is positive and the force push the particle away from the $z$ axis. In the interval $\theta=[0,\pi/2]$, $f_{z}(\theta)$ is negative. For $\theta=[\pi/2,\pi]$, $f_{z}(\theta)$ is positive and the force push the particle away from the $xy$ plane. When $\theta=[\pi/2,\pi-\arccos(1/3)]$, $f_{\rho}(\theta)$ and $f_{z}(\theta)$ are positive and we have a repulsive force between the atoms.

%%%%%%%%%%%%%%%%%%%%%%%%%%%%
%%%%%%%%%%%%%%%%%%%%%%%%%%%%

%%%%%%%%%%%%%%%%%%%%%%%%%%%%
%%%%%%%%%%%%%%%%%%%%%%%%%%%%

It is interesting to compare the interaction energy (\ref{Eint2H}) with the Van der Waals energy between two hydrogen atoms in the ground state. The non-dispersive Van der Waals force is, approximately, \cite{Cohen}
%%%%%%%%%
\begin{equation}
\label{VDW}
{\bf F}_{VW}\cong-36e^{2}\frac{a_{0}^{5}}{|{\bf r}|^{7}}{\hat r}\ .
\end{equation}
%%%%%%%%%

From (\ref{numD}) and (\ref{2HYA7}), the force (\ref{2HYA18}) can be written as
%%%%%%%%
\begin{eqnarray}
\label{Fquadrupolo}
{\bf F}^{Q}\cong 4.45\frac{e^{2}a_{0}^{6}{\bf v}^{2}}{|{\bf r}|^{8}}\left\{\left[1-9\cos^{2}(\theta)\right]{\hat r}-2\cos(\theta)\hat{v}\right\}\ .
\end{eqnarray}
%%%%%%%%%

For distances much higher than the Bohr radius, $a_{0}$, where the results (\ref{Fquadrupolo}) and (\ref{VDW}) are valid, and using the overestimated value (\ref{estimativa}) for $|{\bf v}|$, one can see that the force (\ref{2HYA18}) is negligible in comparison with the Van der Waals force (\ref{VDW}).

The torque (\ref{torque}) vanishes when the vector distance between the atoms, ${\bf r}$, is parallel, anti-parallel or perpendicular to the background vector ${\bf v}$, namely, when $\theta=0,\pi,\pi/2$.

%%%%%%%%%
\section{Final Remarks}
\label{concl}
%%%%%%%%%

In this paper we considered the physical effects induced on the ground state of a hydrogen atom by the presence of a background field in a Lorentz symmetry breaking scenario.

We considered the model studied in reference \cite{BBHepjc2014} in lowest order in the background field. The model exhibits a Lorentz symmetry breaking in the gauge sector controlled by a single background vector. We calculated the energy shift and correction to the wave function for the ground state of the hydrogen atom. We used standard non relativistic perturbation theory.   
We showed that there is no induced electric and magnetic dipole moments on the ground state of the hydrogen atom, but there is an induced four-pole moment, which produces an interaction energy between two hydrogen atoms (both at ground state) placed at a distance $\bf r$ apart. This interaction energy leads to an anisotropic force between the atoms, as well as a torque on the distance vector between the atoms, $\bf r$, with respect to the background vector $\bf v$. We compared this force with the (the non dispersive) Van der Waals one and concluded that this last one is always dominating in the regions where they both are not negligible.

The model we considered is restrict. In a more general situation one should take into account other contributions for the the $k_{F}$ tensor as well as contributions of the $k_{AF}$ tensor, as defined in reference \cite{PRD116002}. To put these results in an experimental context, one would take into account the rotation of the background vector with respect to the laboratory frame \cite{PRL82,arXiv1506.01706}. The effects of this rotation shall not modify the order of magnitude of the results, that are completely out of reach for any spectroscopic experiment nowadays.

\

%%%%%%%%%
\noindent {\bf Acknowledgments}

L.H.C. Borges thanks to FAPESP, under the process 2013/01231-6 for financial support. F.A. Barone thanks to CNPq for financial support. The authors would like to thank J.A. Helay\"el-Neto for reading the paper.
%%%%%%%%%

%%%%%%%%%

%%%%%%%%%

\end{document}